# Improving performance of an analog electronic device using quantum error correction

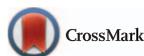

# Improving performance of an analog electronic device using quantum error correction

Corey Ostrove, Brian La Cour , Andrew Lanham and Granville Ott

Applied Research Laboratories, The University of Texas at Austin, P.O. Box 8029, Austin, TX 78713-8029, United States of America

E-mail: blacour@arlut.utexas.edu

## Abstract

The use of analog classical systems for computation is generally thought to be a difficult proposition due to the susceptibility of these devices to noise and the lack of a clear framework for achieving fault-tolerance. We present experimental results for the application of quantum error correction (QEC) techniques to a prototype analog computational device called a quantum emulation device. It is shown that for the gates tested (transversal $Z$, $X$ and $SH$) there is a marked improvement in the performance characteristics of the gate operations following error correction using the 5-Qubit Perfect code. In the case of the $Z$ gate, the median fidelity improved from 0.995 to 0.999 98, a reduction in the gate error by over two orders of magnitude. Other transverse gates similarly show strong improvements.

## 1. Introduction

Quantum information processing techniques provide an optimistic path towards the development of computational devices that can address problems for which no known efficient classical approaches are known to exist. Underlying the usefulness of quantum systems for information processing is the ability to scale up what are now small laboratory scale prototype devices to the sizes necessary for addressing problems that cannot already be tackled with classical hardware. Before the discovery of techniques for performing quantum error-correction (QEC), it was thought that the inherent fragility of quantum coherent systems would make scaling up to these sizes virtually impossible in practice. This was not unlike the situation in the early days of digital computing, when real-time error correction techniques were first contemplated [1]. Beginning with the discovery of Shor's famous 9-qubit code in 1994, and followed by subsequent advances within the field of QEC a path towards large-scale, fault-tolerant quantum computers now exists [2, 3].

Experimental implementations of many of the components that go into the performance of QEC protocols have been tested for most of the potential candidate platforms for quantum computing. Ion-trap quantum computing experiments have been performed that tested the performance of a three-qubit stabilizer code using a system comprised of the hyperfine levels of three trapped beryllium atoms with intentionally induced errors and found a reduction in the gate error by as much as 0.3 for large error probabilities [4]. Further ion-trap experiments have looked at multiple rounds of QEC using the three-qubit phase-flip code on a system of three calcium ions under the influence of intentionally induced errors between each round, which found a degradation in fidelity consistent with first-order insensitivity to induced errors [5]. Experiments performed using logical qubit states encoded as 'cat-states' of a superconducting quantum oscillator reported a 10% enhancement in coherence time using QEC over the longest lived physical qubit comprising the composite system and a factor of 20 times longer than the shortest lived component (a transmon qubit) [6]. A demonstration of a universal set of one-qubit gate operations acting on logically encoded qubit states in a superconducting quantum oscillator has shown an average fidelity for logical gate operations of 0.985 [7]. For superconducting transmon qubits experimental tests have verified first-order insensitivity to induced phase-flip and bit-flip errors using three-qubit repetition codes; however those same experiments also found that for low error probability the overhead of QEC actually reduced overall performance [8]. Experiments on transmons have also demonstrated the implementation of stabilizer measurements of the type necessary for the fault-tolerant implementation of QEC with stabilizer codes [9]. Recent experiments on transmon qubits have also demonstrated a reduction in the failure rate of input state retrieval by as much as a factor of 8.5 using multiple rounds of QEC as compared to rates for unencoded qubits [10].

In nuclear magnetic resonance (NMR) systems, experimental tests of logical operations on a system of five nuclear spins encoded in the 5-qubit Perfect code in the presence of induced errors found that the average gate performance was enhanced relative to what would be expected for an unencoded qubit subject to the same noise [11]. Furthermore, experiments on NMR systems have also demonstrated an improvement using QEC for a system subjected to phase noise and encoded in a 3-qubit phase-flip repetition code [12]. In nitrogen vacancy center qubits, an experimental implementation of the 3-qubit phase-flip code under the influence of induced noise found a reduction in error for large induced-error probabilities [13]. Finally, in photonic systems, tests have been performed using both single-photon and continuous-variable (CV) settings. In the CV setting, an experimental test of a CV generalization of the Shor 9-qubit code in the presence of induced noise found an enhancement using QEC for each of the induced error modes [14]. In the single-photon setting experimental tests of a simple code for protection from coherence loss from spurious $Z$ basis measurements was found to recover the original state following intentionally induced measurements with a fidelity of 0.98 [15]. Additionally, an experiment using a 4-qubit code designed to protect from photon loss errors in a single photon system recovered from induced photon loss errors with a fidelity of 0.80 [16].

A number of analogies are often drawn between quantum mechanical systems and classical analog systems [17]. A key distinction made with regard to computing is that, under certain assumptions, a scalable quantum computer is capable of satisfying the threshold theorem for fault tolerance [18]. In previous work we have shown how one may explicitly embed the same Hilbert space structure found in a gate-based quantum computer directly into a signal processing framework in which information is represented using complex, basebanded analog voltage signals [19]. This analogy allows for a rich set of connections to be drawn between problems in both signal processing and quantum computing applications. In a separate work, for example, we have shown how this mathematical connection can be leveraged to incorporate techniques developed for QEC to solve problems in digital wireless communications [20]. A prototype quantum emulation device (QED) that utilizes this embedding and physically performs the operations described in hardware has been developed and tested [21]. A natural question to ask given this device is whether it is possible to utilize techniques from QEC to enhance the performance of what is otherwise a purely classical analog device. On a practical level this is an interesting question for multiple reasons. Firstly it is believed that error correction in analog devices is very difficult, and the difficulty of this problem has long stymied advances in technologies that rely on analog data processing and collection [22]. Secondly, it is not clear that QEC should work on a classical analog device at all. Regardless of our embedding scheme, the device dynamics are governed purely by classical physics and, as such, the assumptions inherent in QEC's formulation, such a linearity of errors, are not necessarily justified. In section 2 we discuss a simple example (additive white Gaussian noise) where it is possible to directly express the classical errors in the quantum formalism. In this work we demonstrate that in practice, despite the above caveats, QEC protocols do in fact provide additional robustness to noise in our system and improve its performance overall.

## 2. Modeling classical errors as quantum operations

In order to effectively make use of QEC techniques in our alternative setting we first need to recast the modeling of our system's error dynamics into the same mathematical framework used in the quantum mechanical setting. Quantum operations provide one such framework and are used to model the evolution of noisy quantum system dynamics. The quantum operations formalism provides the tools needed to describe the evolution of open quantum systems, those coupled to external environmental degrees of freedom, along with the apparent non-unitary system evolution that can occur. The mapping $\rho \mapsto \mathcal{E}(\rho)$ is a superoperator that maps the input density operator $\rho$ acting on a Hilbert space $\mathcal{H}$ to the final density operator $\mathcal{E}(\rho)$, here taken to be also acting on $\mathcal{H}$, and is called the quantum operation or, for trace-preserving maps, the quantum channel [23].

Quantum operations can be described using the operator-sum representation in which the evolution of the system is specified by a discrete set $\{E_k\}$ of operators on the Hilbert space, called the Kraus operators [24]. In this formulation, the quantum operation takes the form

$$\mathcal{E}(\rho) = \sum_k E_k \rho E_k^\dagger. \tag{1}$$

where, for a trace-preserving quantum channel, $\sum_k E_k E_k^\dagger = I$ is the identity.

Operators from the Pauli group are an example of Kraus operators. The Pauli group for a system of $n$ qubits is given by the set of all $n$-fold tensor products of the one-qubit Pauli matrices $\sigma_0 = I, \sigma_1 = X, \sigma_2 = Y, \sigma_3 = Z$. The Pauli group on $n$ qubits forms a complete basis for $2^n \times 2^n$ matrices, so it is always possible to rewrite the operator-sum representation of a quantum operation in a canonical form by rewriting the operation elements as linear combinations of Pauli group elements [23].

The depolarizing channel is a prototypical channel within the QEC literature. Correcting depolarizing errors on a number of qubits is as hard as correcting arbitrary errors on those qubits and, so, it is a simple and useful stand-in [23]. For the purposes of error correction in classical analog systems, the depolarizing channel is of particular interest because it can be shown to directly correspond to the presence of additive Gaussian white noise (AWGN) in a system [25]. The depolarizing channel is a noise process in which with some probability $p$ we lose all information about the state of our system and it is replaced with the maximally mixed state $I/N$ where $N = 2^n$ is the dimension of the system. As a quantum operation we can write the generalized depolarizing channel as follows:

$$\mathcal{E}(\rho) = (1-p)\rho + \frac{p}{N}I. \tag{2}$$

Making use of the identity

$$I = \frac{1}{N} \sum_{m_1=0}^{3} \cdots \sum_{m_n=0}^{3} \sigma_{m_1} \otimes \cdots \otimes \sigma_{m_n} \rho \, \sigma_{m_1} \otimes \cdots \otimes \sigma_{m_n}, \tag{3}$$

the depolarizing channel can be written in the operator-sum representation. For the one-qubit case, this is given by

$$\mathcal{E}(\rho) = (1-p')I\rho I + \frac{p'}{3}(X\rho X + Y\rho Y + Z\rho Z), \tag{4}$$

where $p' = 3p/4$.

To see the connection to AWGN suppose the quantum state $|\psi\rangle$ is represented by the time-domain signal $\psi$ given by

$$\psi(t) = \sum_{x=0}^{N-1} \phi_x(t)\langle x|\psi\rangle = \sum_{x=0}^{N-1} \langle x|\psi\rangle \exp\left[i\sum_{k=1}^{n}(-1)^{x_k}\omega_k t\right], \tag{5}$$

where $x = x_{n-1}2^{n-1} + \cdots + x_0 2^0$ and $\omega_k = 2^k \omega_0$ for some $\omega_0 > 0$. Additive noise produces a stochastic signal $\tilde{\psi} = \psi + w$, where $w$ is a zero-mean complex white Gaussian noise process with spectral density $\sigma^2$ such that $\mathsf{E}[w^*(t)w(t')] = \sigma^2 \delta(t-t')$, where $\mathsf{E}$ represents an expectation value. Note that $\tilde{\psi}$ is outside the Hilbert space of $\psi$ since $w$ is a broadband signal. Projecting $\tilde{\psi}$ back into this space, which is done by narrowband filtering, yields a quantum state of the form $|\tilde{\psi}\rangle \propto |\psi\rangle + |\nu\rangle$. Here, $|\nu\rangle$ is represented by a stochastic signal given by

$$\nu(t) = \sum_{x=0}^{N-1} \phi_x(t)\langle x|w\rangle = \sqrt{\frac{\sigma^2}{T}} \sum_{x=0}^{N-1} z_x \phi_x(t), \tag{6}$$

where $T$ is a multiple of $2\pi/\omega_0$, $\{z_x\}_x$ are independent standard complex Gaussian random variables, and we have used the inner product definition

$$\langle x|w\rangle = \frac{1}{T}\int_0^T \phi_x(t)^* w(t)\, dt. \tag{7}$$

The corresponding quantum channel may be found by taking the expectation value of the outer product $|\tilde{\psi}\rangle\langle\tilde{\psi}|$. Since

$$\mathsf{E}[\langle x|\tilde{\psi}\rangle\langle\tilde{\psi}|x'\rangle] = \langle x|\psi\rangle\langle\psi|x'\rangle + \frac{\sigma^2}{T}\mathsf{E}[z_x z_{x'}^*] = \langle x|\psi\rangle\langle\psi|x'\rangle + \frac{\sigma^2}{T}\delta_{xx'}, \tag{8}$$

we deduce that the quantum channel is given by

$$\mathcal{E}(\rho) = \left(1 + \frac{N\sigma^2}{T}\right)^{-1}\left[\rho + \frac{\sigma^2}{T}I\right]. \tag{9}$$

which is of the same form as equation (2). This shows how a classical noise process, in this case additive white Gaussian noise, can be described by an equivalent quantum operation.

## 3. Description of the QEC protocol

For our experiment we used the 5-qubit Perfect code, which is the smallest code capable of correcting an arbitrary error on a single physical qubit [26]. The 5-qubit code is a stabilizer code with the generators given in table 1.

To encode the logical single-qubit state $|\psi\rangle = \alpha|0\rangle + \beta|1\rangle$ into the physical state $|\bar{\psi}\rangle = \alpha|\bar{0}\rangle + \beta|\bar{1}\rangle$, we use the code words

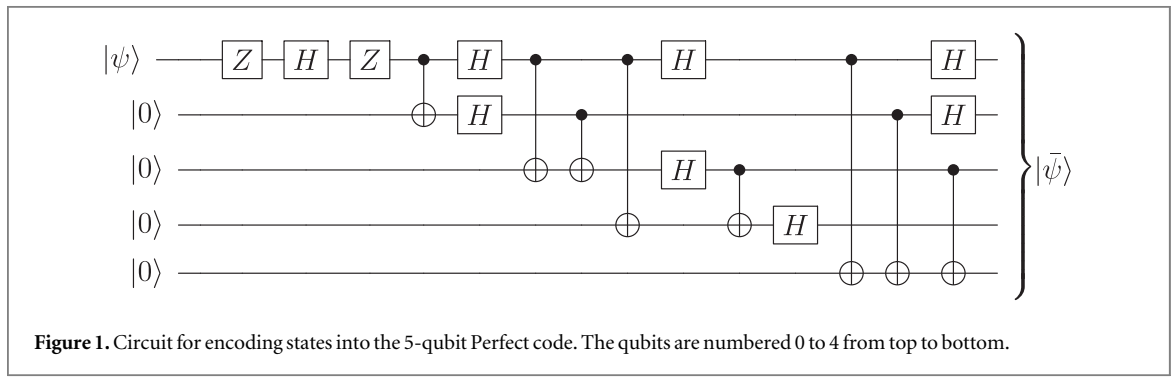

**Figure 1.** Circuit for encoding states into the 5-qubit Perfect code. The qubits are numbered 0 to 4 from top to bottom.

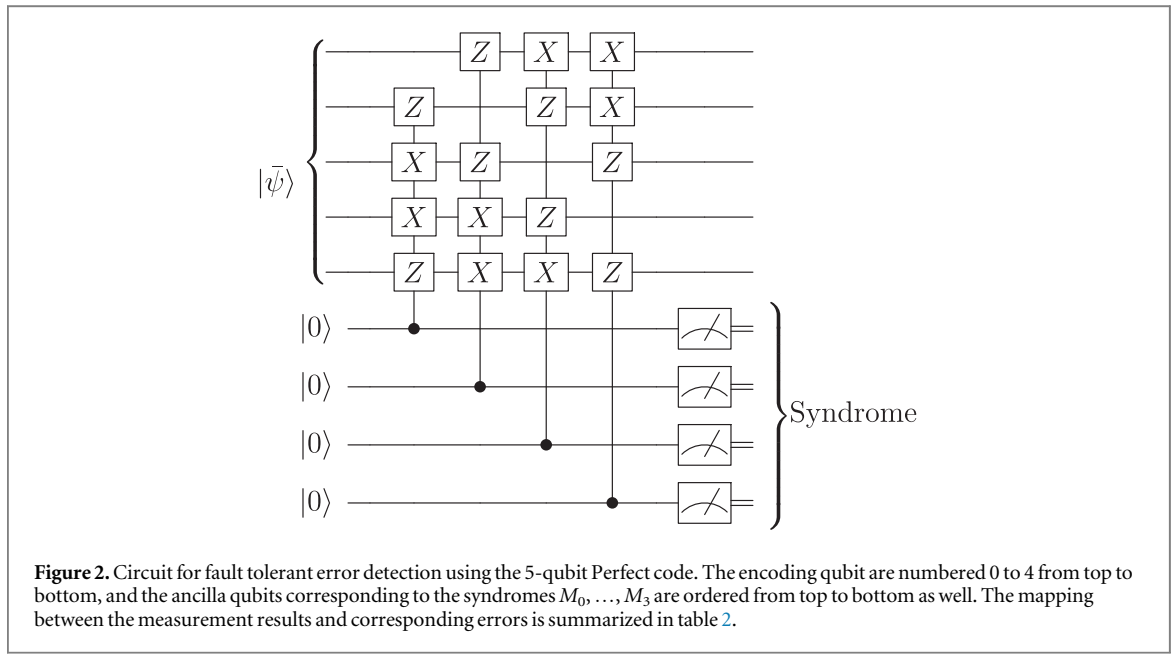

**Figure 2.** Circuit for fault tolerant error detection using the 5-qubit Perfect code. The encoding qubit are numbered 0 to 4 from top to bottom, and the ancilla qubits corresponding to the syndromes $M_0, \ldots, M_3$ are ordered from top to bottom as well. The mapping between the measurement results and corresponding errors is summarized in table 2.

**Table 1.** Generators for the stabilizer group of the 5-qubit Perfect code.

$M_0 = Z_1 X_2 X_3 Z_4$
$M_1 = Z_0 Z_2 X_3 X_4$
$M_2 = X_0 Z_1 Z_3 X_4$
$M_3 = X_0 X_1 Z_2 Z_4$

$$|\bar{0}\rangle = \frac{1}{4}(I + M_0)(I + M_1)(I + M_2)(I + M_3)|00000\rangle \qquad (10a)$$

$$|\bar{1}\rangle = \frac{1}{4}(I + M_0)(I + M_1)(I + M_2)(I + M_3)|11111\rangle, \qquad (10b)$$

where a little endian qubit numbering convention (0 to 4, read right to left) is adopted. The encoding circuit is given in figure 1.

In addition to the encoding circuit, we also need the corresponding syndrome detection circuit. To perform syndrome detection we use a fault-tolerant circuit construction introduced by DiVincenzo and in the form presented by Mermin [2, 27]. The circuit diagram for this is given in figure 2. This circuit construction, which uses four additional ancillary qubits, is not the most qubit-resource efficient fault-tolerant syndrome detection scheme but is easy to implement and understand. (For a more resource-efficient scheme, see [28, 29].) table 2 gives the correspondence between the measurement results on the ancilla qubits in figure 2 and the error that was measured. Each controlled operation in figure 2 projects the system into the ±1 eigenspace of each of the stabilizer group generators, and the measurement on the corresponding ancilla tells into which of the two eigenspaces the state was projected. Since all of the Pauli operators are involutions, the correction operation is simply to apply the same Pauli operation indicated by the syndrome measurement.

**Table 2.** Syndrome-Error Correspondence for 5-Qubit Code.

| Syndrome $M_0M_1M_2M_3$ | Error | Syndrome $M_0M_1M_2M_3$ | Error |
|---|---|---|---|
| 0000 | No Error | 1000 | $Z_2$ |
| 0001 | $Z_1$ | 1001 | $X_4$ |
| 0010 | $X_3$ | 1010 | $X_1$ |
| 0011 | $Z_0$ | 1011 | $Y_1$ |
| 0100 | $X_0$ | 1100 | $Z_3$ |
| 0101 | $X_2$ | 1101 | $Y_2$ |
| 0110 | $Z_4$ | 1110 | $Y_3$ |
| 0111 | $Y_0$ | 1111 | $Y_4$ |

Transverse (i.e., separable) gates are needed in order to implement fault-tolerant encoded gates on the encoded states. The 5-qubit Perfect code has transversal Pauli gates, with $\bar{\sigma}_m = \sigma_m^{\otimes 5}$, as well as a set of Clifford operations given by

$$K_{s_x,s_y,s_z} = \exp\left[\frac{i\pi}{3\sqrt{3}}(s_x X + s_y Y + s_z Z)\right] \text{for} s_x, s_y, s_z \in \{-,+\} \quad (11)$$

that are also transversal, with $\bar{K}_{s_x,s_y,s_z} = K_{s_x,s_y,s_z}^{\otimes 5}$ [30]. This gives an easy way to implement a set of test gates with which to evaluate improved performance on our device. In the performance experiments described in section 4, the system is benchmarked using the Pauli $Z$ and $X$ gates as well as the Clifford $SH$ gate. The $SH$ gate is given by the product of the phase gate $S = \sqrt{Z}$ and the Hadamard gate $H = (X + Z)/\sqrt{(2)}$ and corresponds to $K_{+,+,+}$ in equation (11). Thus,

$$K_{+,+,+} = e^{i\pi/4}SH = \frac{1}{2}\begin{pmatrix} 1+i & 1+i \\ -1+i & 1-i \end{pmatrix}. \quad (12)$$

## 4. Experiments

To evaluate the performance of our hardware device we performed a series of tests to determine performance with and without the use of QEC protocols. We then compared the single-gate fidelity in each of the two cases using a random set of logical input states.

### 4.1. Experimental design

The details of the hardware are described elsewhere [21]. Two of the five encoding qubits were represented in the frequency domain using signals with four narrowband tonals at $\pm 1000$ Hz and $\pm 3000$ Hz. The other three qubits were represented in the time domain using a wavetrain of eight such signals. (The classical signal representation requires time-frequency resources that will, of course, scale exponentially with the number of qubits.) Gate operations on frequency domain qubits were performed in hardware using analog filters, operational amplifiers, and four-quadrant multipliers. Likewise, the signal multiplication operations necessary for performing gates on the time-domain qubits was performed in hardware with the reordering operations handled digitally following an analog-to-digital converter (ADC). The configuration was chosen such that qubits 0 and 1 were represented in the frequency domain, while qubits 2 through 4 were represented in the time domain. The choice was arbitrary, and other configurations give similar results. For more information regarding the time domain encoding scheme and the corresponding gate operations see [31]. The general workflow of the experiments is detailed in the flowchart in figure 3.

The first stage of each of the experiments is a software pre-processing step in which we generate a set of 100 pure state inputs uniformly at random according to the Haar measure [23]. Each of these states is synthesized digitally and, in software, encoded into the 5-qubit code using the circuit given in figure 1. After this pre-processing stage, the physical analog signals are generated with a digital-to-analog converter (DAC), and a selected transversal gate is applied in hardware using analog electronics. The transformed signals are then sampled digitally using an ADC and buffered in memory. This digitized signal is converted back to the corresponding quantum state, upon which syndrome measurements may be performed using measurements based on the Born rule. Finally, a software-based post-processing stage occurs in which the transformed states have the syndrome detection circuit of figure 2 applied, along with the appropriate correction gate based on the syndrome measurement results as given in table 2. The decoding circuit is then applied (The circuit given in figure 1 but in reverse order) and the fidelity between the measured output state and the correct result is calculated. Note that only the encoded gate is performed in hardware—all encoding, error correction, and decoding is performed in software.

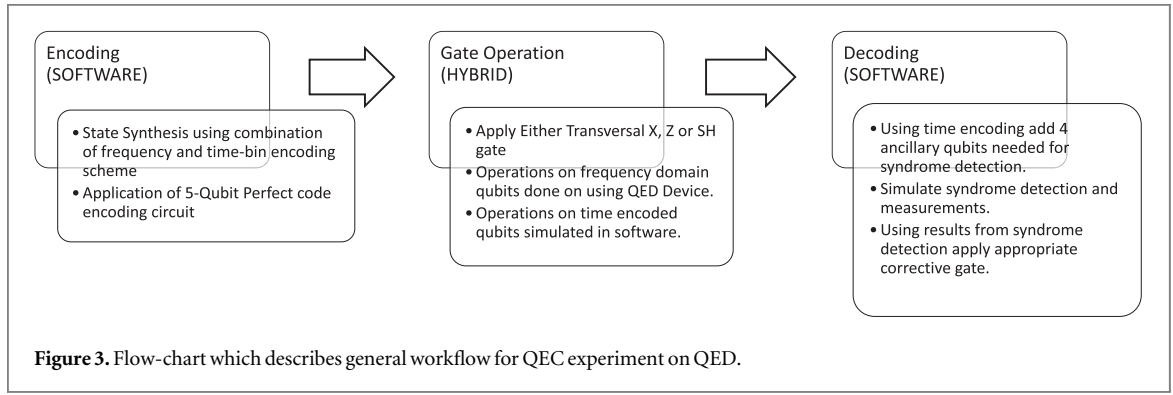

**Figure 3.** Flow-chart which describes general workflow for QEC experiment on QED.

In addition to the encoded and error-corrected gate operations, a set of control runs was performed for each of the gates using the same set of input states and the same number of repetitions. This control run was performed directly on the one-qubit input states encoded in the frequency domain rather than on the logically encoded states.

### 4.2. Performance metrics
A common metric used to quantify the performance of a quantum gate is the gate fidelity. Let $|\psi\rangle$ denote the notional input state for our gate, and let $U$ denote the gate that we intend to apply. In practice what is actually implemented is a noisy version of $U$, which may be described by a quantum operation $\mathcal{E}$. The gate fidelity for a particular input state, defined as

$$F = \sqrt{\langle\psi|\ U^\dagger\ \mathcal{E}(|\psi\rangle\langle\psi|)\ U|\psi\rangle}, \quad (13)$$

measures how closely the noisy implementation of $U$ approximates the desired one. A more general (input-independent) measure of the performance is the median gate fidelity, denoted $\bar{F}$, which is taken over an ensemble random input states and, for each one, several repetitions of a given gate operation. A median is preferred over the mean in order to characterize typical behavior in highly skewed data. Since fidelity is bounded above by 1, changes in the fidelity as a result of error-correction can be very small in absolute terms The performance of a system in the long-term often has an exponential sensitivity to the infidelity $(1 - F)$ of the operations and, so, we define a metric that reflects this sensitivity, which we call the log-fidelity, denoted $f$, and define it as

$$f = -\log_{10}(1 - F). \quad (14)$$

The choice of base 10 for the logarithm in equation (14) gives the log fidelity a simple interpretation in terms of a more common colloquial measure, the number of nines in the fidelity. As defined, the log-fidelity is equal to the number of nines plus an interpolation between an integer number of nines. We note that the definition of this measure draws analogy between the log-fidelity and the decibel scale used in classical systems for characterizing performance in terms of the signal-to-noise ratio (SNR). Use of the median fidelity also has the desirable property that median log-fidelity, $\bar{f}$, is equal to $-\log_{10}(1 - \bar{F})$, a property not shared by the mean fidelity.

### 4.3. Results
Starting with the $Z$ gate we find that without QEC implemented, the median fidelity over the entire set of 100 input state randomizations and 1000 experimental runs was 0.994 63 with an approximate 95% confidence interval of (0.994 60, 0.994 67) obtained by bootstrapping over $10^4$ random samples. Also useful for visualization of the performance statistics is the cumulative distribution function (cdf), which allows one to more easily compare the spread of the distributions. The results for the $Z$ gate with QEC implemented tell a more interesting story. In figure 4 we compare the cdf and corresponding probability density functions (pdf) for the corrected and uncorrected Z gates. After error correction, it was found that the median fidelity for the $Z$ gate increased to 0.999 976 4 (0.999 976 1, 0.999 976 8). This was true despite the fact that the experimental error correction procedure tended to broaden the tail on the fidelity distribution, resulting in several outliers. This can be observed in the appearance of a bimodal clustering of low fidelity results that can be seen in the inset plot of figure 4. The reasons behind this tail broadening behavior will be expanded on further in section 5. Looking at the log-fidelities given in figure 4 shows, however, that despite the fact that QEC has a tendency to broaden the tails by periodically causing low fidelity outcomes, most of the time the procedure substantially improves the performance. This is further evidenced by looking at the change in the median log-fidelity, where we find, without error correction, the median log-fidelity for the *Z* gate is 2.270 (2.267, 2.273) and, with error-correction, it improves to 4.628 (4.622, 4.634).

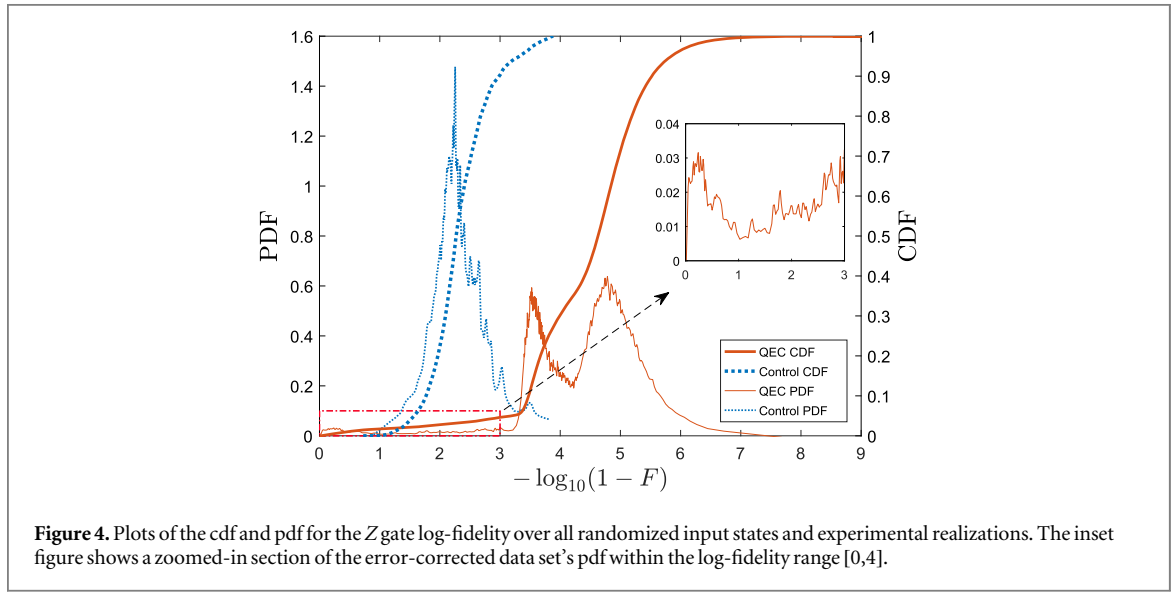

**Figure 4.** Plots of the cdf and pdf for the *Z* gate log-fidelity over all randomized input states and experimental realizations. The inset figure shows a zoomed-in section of the error-corrected data set's pdf within the log-fidelity range [0,4].

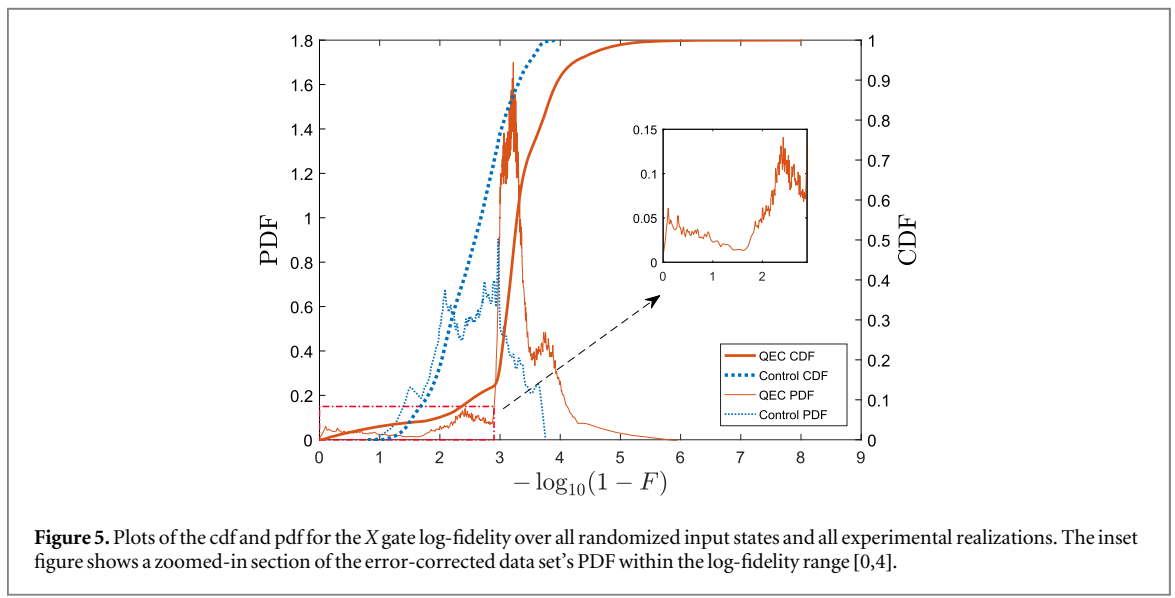

**Figure 5.** Plots of the cdf and pdf for the *X* gate log-fidelity over all randomized input states and all experimental realizations. The inset figure shows a zoomed-in section of the error-corrected data set's PDF within the log-fidelity range [0,4].

The *X* and *SH* gate results are similar to those found for the *Z* gate, and a summary of the results can be found in figures 5 and 6 respectively. In figure 5 it can be seen in both cases that we again have a tail broadening effect in the error-corrected results with a similar bimodal clustering of the low-fidelity outcomes. However, the vast majority of events demonstrate a substantial improvement. For the *X* gate, despite the tail broadening, the median fidelity after error correction, 0.999 426 (0.999 423, 0.999 429), was significantly higher than that of the uncorrected gate, which was found to be 0.997 39 (0.997 36, 0.997 43). The median log-fidelity after QEC was found to be 3.241 (3.239, 3.243), as compared to 2.584 (2.578, 2.590) without error correction. Likewise, for the *SH* gate, the median fidelity of the uncorrected *SH* gate was found to be 0.995 27 (0.995 24, 0.995 30), which was improved to 0.999 905 5 (0.999 904 8, 0.999 906 1) with QEC. The median log-fidelity of the uncorrected *SH* gate was found to be 2.325 (2.322, 2.328); after QEC this increased to 4.024 (4.022, 4.027).

## 5. Discussion

One of the key issues raised by the experimental results with the application of QEC to the QED is the tail broadening effect that is observed on the fidelity. In order for QEC to be useful it is important to understand the conditions and noise processes which contribute this behavior. The device itself is subject to a myriad of classical noise sources including everything from Johnson noise, inherent to finite-temperature operation, to phase and frequency drift resulting from signal filtering in the gate and measurement operations. For the purposes of higher level modeling, however, it suffices to conceptually model the errors in the system as resulting from two main types of error. The first error source is gate errors caused by the imperfect implementation of the gate operations, and the second is due to general imperfections in the circuit construction, which puts a floor on the fidelity of our state representation even in the absence of any gate operation. Gate operations on the QED are implemented using analog multiplication operations with gate coefficients defined by a corresponding set of analog DC voltage values. The DC values that define the coefficients are inherently imperfect, however. Due to a combination of quantization error caused by the finite resolution of the digital-to-analog converters (DACs) and stochastic noise sources such as thermal noise, these DC values will fluctuate and be randomly distributed about the desired values. We suspect that different gates will have different fidelities due to the differing DC voltages needed to realize each in the hardware implementation. The Z gate, for example, has similiar DC voltages to an identity gate and would therefore be expected to perform better.

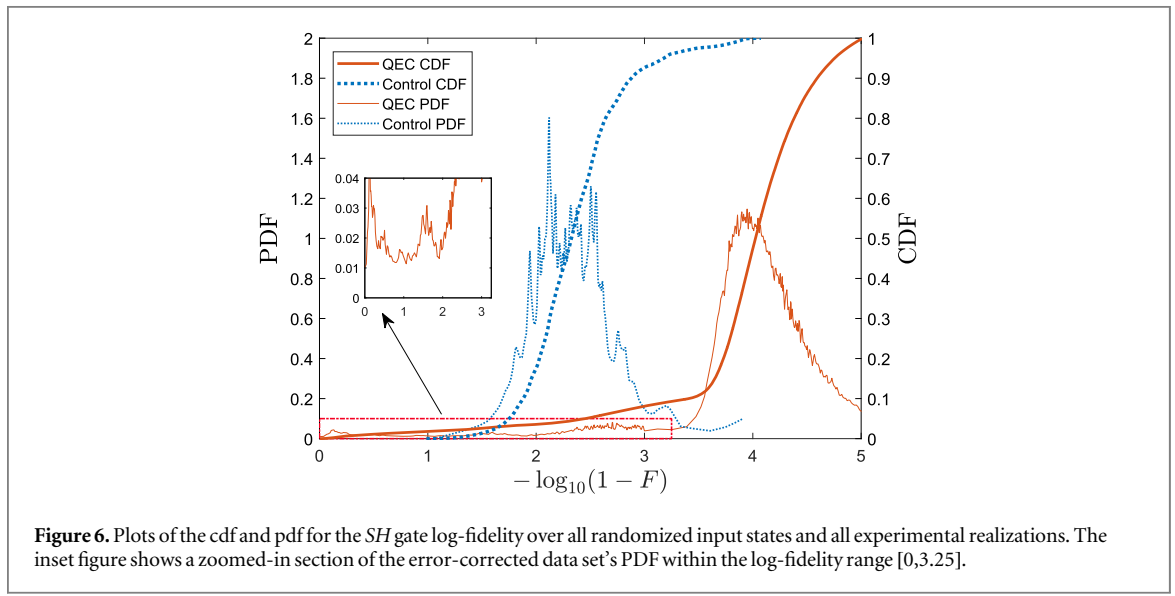

**Figure 6.** Plots of the cdf and pdf for the *SH* gate log-fidelity over all randomized input states and all experimental realizations. The inset figure shows a zoomed-in section of the error-corrected data set's PDF within the log-fidelity range [0,3.25].

A key feature of all of the above noise processes is that they are inherently continuous, whereas the QEC corrections we apply are inherently discrete. In particular, codes such as the 5-qubit perfect code are constructed with a model in mind that is based on the idea that errors act on qubits within the state locally and independently. In cases where too many qubits are hit by the noise process it is possible to misidentify the error syndrome and in the correction process transform the state of the system into one nearly orthogonal to the original. From this we can see a likely candidate for the source of the bimodal clustering of low fidelity values in the experimental results presented in section 4.3. We can see a clear example of this in figure 4, where there are two modes of the QEC pdf at log-fidelities of about 3.5 and 5. This leaves open an interesting possibility for improving the performance of the QEC protocols in the device by designing decoding procedures which leverage the additional information we have access to from having an explicit representation of the state of the system and which performs corrections in a continuous manner. We explore this idea more directly in the context of applying QEC to wireless communication applications, and it is likely the ideas developed in that context would be similarly applicable to the QED device [20].

## 6. Conclusions

We have shown that the techniques of quantum error correction can be successfully applied in domains seemingly far removed from standard quantum mechanical systems. The QED device implements a classical representation of quantum states based on pairs of analog voltage signals to perform its information processing. As a classical device it is in principle subject to a whole myriad of errors, some of which, as in the case of AWGN, are representable as quantum operations. Yet, in practice, it is found that QEC nonetheless yields a practical performance boost. All three gates studied—the *X*, *Z* and *SH* gates—showed marked improvement in their operating characteristics, with an average increase in log-fidelity over the three gates of about 1.57. Given the general effectiveness of QEC at improving the device's performance it seems reasonable to suppose that in the above experiments the device is operating in a regime in which the errors are dominated by those in which there is a quantum analogue amenable to QEC, even if our models have not fully captured the details of those analogues yet. The results of this work may have applicability beyond computing and could serve as a basis for advanced techniques in robust, fault-tolerant classical communication through the use of quantum error correction protocols on classical messages.

## Acknowledgments

This work was supported by the Office of Naval Research under Grant Nos. N00014-14-1-0323 and N00014-17-1-2107.

## ORCID iDs

Brian La Cour  https://orcid.org/0000-0001-7899-0938